\def\year{2022}\relax
\documentclass[letterpaper]{article} 
\usepackage{aaai24}
\usepackage{times}  
\usepackage{helvet}  
\usepackage{courier}  
\usepackage[hyphens]{url}  
\usepackage{graphicx} 
\usepackage{natbib}  
\usepackage{caption} 
\DeclareCaptionStyle{ruled}{labelfont=normalfont,labelsep=colon,strut=off} 
\usepackage{algorithm}
\usepackage{algorithmic}
\usepackage{csquotes}
\usepackage{booktabs}
\usepackage{subcaption}

\usepackage{newfloat}
\usepackage{listings}
\floatstyle{ruled}
\newfloat{listing}{tb}{lst}{}
\floatname{listing}{Listing}

\usepackage{xcolor}
\newcommand{\answerYes}[1]{\textcolor{blue}{#1}} 
 
\newcommand{\answerNA}[1]{\textcolor{gray}{#1}}

\newcommand{\ignore}[1]{{}}

\nocopyright
\title{InstaSynth: Opportunities and Challenges in Generating Synthetic Instagram Data with chatGPT for Sponsored Content Detection\footnote{*To appear at the 18th International AAAI Conference on Web and Social
Media (ICWSM 2024) -- please cite accordingly.}}

\author{
      Thales Bertaglia,\textsuperscript{\rm 1}
      Lily Heisig,\textsuperscript{\rm 2}
     Rishabh Kaushal,\textsuperscript{\rm 3,\rm 4}
      Adriana Iamnitchi,\textsuperscript{\rm 3}
}
\affiliations{

     \textsuperscript{\rm 1} Utrecht University, The Netherlands\\
     \textsuperscript{\rm 2} University College Maastricht, Maastricht University, The Netherlands\\
     \textsuperscript{\rm 3} Institute of Data Science, Department of Advanced Computing Sciences, Maastricht University, The Netherlands\\
     \textsuperscript{\rm 4} Department of Information Technology, Indira Gandhi Delhi Technical University for Women, India\\
   t.f.costabertaglia@uu.nl, l.heisig@student.maastrichtuniversity.nl, \{rishabh.kaushal, a.iamnitchi\}@maastrichtuniversity.nl}

\begin{document}

\maketitle

\begin{abstract}
Large Language Models (LLMs) raise concerns about lowering the cost of generating texts that could be used for unethical or illegal purposes, especially on social media. This paper investigates the promise of such models to help enforce legal requirements related to the disclosure of sponsored content online. 
We investigate the use of LLMs for generating synthetic Instagram captions 
with two objectives: The first objective (\textit{fidelity}) is to produce realistic synthetic datasets. For this, we implement content-level and network-level metrics to assess whether synthetic captions are realistic. 
The second objective (\textit{utility}) is to create synthetic data useful for sponsored content detection. 
For this, we evaluate the effectiveness of the generated synthetic data for training classifiers to identify undisclosed advertisements on Instagram.
Our investigations show that the objectives of fidelity and utility may conflict and that prompt engineering is a useful but insufficient strategy. Additionally, we find that while individual synthetic posts may appear realistic, collectively they lack diversity, topic connectivity, and realistic user interaction patterns. 

\end{abstract}

\section{Introduction}
LLMs like chatGPT present opportunities and challenges for the digital realm, especially in social media. While the potential misuse of LLMs in generating fake news~\cite{caramancion2023harnessing} and fraudulent profiles~\cite{ayoobi2023looming} is evident, they could also benefit data annotation~\cite{tornberg2023chatgpt,kuzman2023chatgpt} and synthetic data generation for myriad text classification tasks, including sentiment analysis and sarcasm detection~\cite{dai2023auggpt,moller2023prompt,veselovsky2023generating}.

This paper evaluates the promise of using LLMs, and specifically chatGPT (\textit{gpt-3.5-turbo}), to generate synthetic captions for Instagram posts. As a case study, we focus on the task of sponsored content detection (SCD) on Instagram, aiming to identify undisclosed ads. Instagram is one of the leading platforms for influencer marketing and, thus, for publishing sponsored posts. The proper disclosure of ads is a legal requirement. Therefore, this task is fundamental for guaranteeing transparency in social media, protecting consumers from misleading or harmful ads, and ensuring legal compliance~\cite{fatima2023deploying,shen2022your,zarei2020characterising, bertaglia2023closing, ershovEffectsInfluencerAdvertising2020}. 
Yet, due to limited API access and costly data annotation, data for developing machine learning solutions for detecting such content is scarce. Moreover, undisclosed ads are inherently challenging to identify and, therefore, collect~\cite{bertaglia2023closing}. Generating faithful synthetic data has the potential to mitigate this issue. Our central research question is: \emph{How well can chatGPT 3.5 bridge this gap by generating synthetic datasets that are realistic enough to address the task of sponsored content detection effectively?} 

While we experiment with one particular version of a particular model during a period of rapid advancements in LLMs, our study yields insights that have broader implications. For example, we observed unrealistic distributions regarding content diversity, representation, and structural connectivity in the synthetic Instagram posts. These observations can guide the evaluation of other LLMs for synthetic data generation. Moreover, in assessing the usefulness of synthetic datasets for identifying undisclosed advertisements, we found conflicting objectives between model effectiveness and the authenticity of the synthetic data compared to real data. This finding indicates that utility and fidelity should be considered separately in evaluations. Lastly, we conclude that these limitations can be addressed more reliably through post-processing in addition to improvements in prompt design. All code and instructions necessary to reproduce our experiments are publicly available on GitHub~\cite{github-repo}.


\section{Related Work}
\label{sec:related_work}

Generating synthetic data is critical in domains with limited resources and critical privacy concerns. Prior research has focused on generating synthetic data in multiple domains, including healthcare~\cite{dahmen2019synsys,dankar2021fake}, IoT~\cite{anderson2014synthetic}, and software engineering~\cite{soltana2017synthetic}.
Before Large Language Models (LLMs), various data augmentation techniques~\cite{liu2020survey,shorten2021text,feng2021survey} have been proposed at character level~\cite{coulombe2018text,saito2017improving} and word level~\cite{wei2019eda,miller1995wordnet,guo2020sequence,qiu2020easyaug}. 

LLM-based data augmentation includes three key aspects: prompt design techniques, evaluation of synthetic data through similarity metrics, and downstream tasks. 
Various strategies employed prompt design to generate the desired outputs from LLMs. The most straightforward approach relies on general instructions such as \enquote{Generate X sentences, ensure diversity}. On a more sophisticated level, \citet{jiao2023chatgpt} introduced the LLM-generated strategy, using the LLM itself to create prompts for specific tasks, for instance \enquote{Provide ten concise prompts for a translation task}.~\citet{veselovsky2023generating,dai2023auggpt,josifoski2023exploiting} show that incorporating examples into the prompt and then instructing the model to generate similar content is effective. In contrast, the concept-teaching (or taxonomy-based) approach introduces conceptual definitions to the prompt by describing labels or setting style directives~\cite{moller2023prompt,ubani2023zeroshotdataaug}. The instructional prompt strategy requires the model to include specific words or features (e.g. dialogue, advertising narration etc.) in the generated content~\cite{eldan2023tinystories}. 





Evaluating the faithfulness of synthetic data generated by LLMs through prompt engineering to real data is fundamental for reliable research. \citet{veselovsky2023generating} note that such data is often \textit{unfaithful}. Related work proposed multiple metrics to measure this similarity. Embedding-based cosine similarity offers a simple method to gauge the closeness between synthetic and real data through their representative embeddings. In a more complex approach, machine learning models (discriminators) are trained to differentiate between the two data types; some examples are the  \textit{believability} model introduced by \citet{veselovsky2023generating} and the propensity or utility scores referenced in various studies~\cite{el2020practical,snoke2018general,dankar2021fake}. \citet{huang2022frustratingly} introduced the TransRate metric, which evaluates the effectiveness of transfer learning by analysing the mutual information between datasets. Content-based metrics focus on distinctive textual features between synthetic and real data, including n-gram overlap~\cite{lim2023artificial,ubani2023zeroshotdataaug},  sentiment analysis, and Flesch readability scores~\cite{flesch1948new}. Lastly, \citet{lim2023artificial} and \citet{hamalainen2023evaluating} use human evaluation to assess the utility of AI-generated content.

The utility of LLMs in data augmentation has been a focal point in recent research, especially regarding their efficacy across diverse downstream tasks. \citet{qin2023chatgpt} explored chatGPT's potential in zero-shot learning across multiple NLP datasets. While \citet{moller2023prompt} and \citet{ubani2023zeroshotdataaug} noticed improvements in specific areas like sentiment analysis and hate speech detection, they found the results less consistent across other tasks. \citet{bang2023multitask} took a broader approach, assessing chatGPT's proficiency in hallucination, reasoning, and interactivity. Some studies highlighted the model's promise in niche applications such as health messaging or text classification~\cite{lim2023artificial,dai2023auggpt}. However, \citet{veselovsky2023generating} emphasised the importance of prompt strategies, noting better performance in sarcasm detection when including examples in the prompt. In contrast, \citet{moller2023prompt} found varying results, with some tasks benefiting from synthetic data and others showing no improvement. 

\section{Methodology}
\label{sec:methodology}

This section outlines our research approach. We discuss the prompt engineering techniques we use for synthetic data generation, describe the metrics used for evaluation, and provide details about the real Instagram datasets we use. 

\paragraph{Prompt Engineering}
Our experiments explored different prompt settings inspired by the strategies discussed in Section~\ref{sec:related_work}. These settings were designed to evaluate the effect of various conditions on the content generated by the model. All strategies included two distinct prompts: one specifically for generating \textit{Sponsored} captions and the other, \textit{Non-Sponsored}. The base prompt explicitly instructs the model to generate a given number of captions for Instagram posts by influencers, ensuring a diverse range in wording, sentence length (falling between 10 and 300 tokens), number of sentences, hashtags, and user tags. We emphasised the generation of original and creative captions, avoiding monotonous or repetitive content. For sponsored posts, we also instructed the model to generate undisclosed ads since this is the focus of the downstream task. We employed four different prompt variations, all built upon the base prompt:

\begin{itemize}
    \item \textbf{Base Prompt}: This setting only provided the basic prompt without any examples.
    \item \textbf{Fixed Examples}: This setting includes five captions from the real dataset to the base prompt as examples. We selected five distinct captions for sponsored posts and another set of five for non-sponsored.
    \item \textbf{Random Examples}: Analogous to Fixed Examples; however, in each request, we included five different (randomly sampled) captions from the real dataset instead of fixed examples.
    \item \textbf{Imitation}: Similar to the Random Examples setting, this used a variety of samples from the real dataset. The distinguishing feature here was the modification of the base prompt. We removed any explicit mentions of Instagram. Instead, the revised prompt instructed the model to generate texts mimicking the examples. This prompt aims to reduce the reliance on the implicit concepts about Instagram encoded into chatGPT, which are not transparent.
\end{itemize}

The full prompts and examples of captions generated by each strategy are available on GitHub~\cite{github-repo}.

\paragraph{Evaluating Synthetic Data}
\label{sec:metrics}

Rigorous evaluation is essential to assess the \textit{fidelity} of the generated captions. Therefore, we employ a diverse set of metrics to analyse different aspects relevant to Instagram captions. We split the metrics into four semantic categories, each focusing on different facets of the content:

\begin{enumerate}
    \item \textbf{Caption Composition Metrics:} Metrics that provide insight into the structure and diversity of the captions. It includes data such as the average and standard deviation of caption lengths, vocabulary size, and the usage of hashtags, user tags, and emojis.
    
    \item \textbf{Content-specific Metrics:} Metrics detailing the overlap between synthetic and real data of n-grams and other relevant features of Instagram captions. It offers insights into the usage of specific phrases and other textual elements.
    
    \item \textbf{Embedding Similarity Metrics:} Metrics that assess the similarity between embeddings of the captions, focusing on the cosine similarity distributions. We use OpenAI's \textit{text-embedding-ada-002} embeddings~\cite{openai-embs}, which output vectors with 1536 dimensions.
    
    \item \textbf{Network Metrics:} Instagram posts often refer to other users or posts via hashtags and user tags, creating a network structure. This structure is fundamental to understanding the platform's dynamic and offers valuable insights into the quality of the captions. We analyse metrics related to network connectivity for three networks, representing the co-occurrence of hashtags, user tags, and the combination of both.
\end{enumerate}

To assess the \textit{utility} of synthetic captions, we use performance metrics related to the classification task of detecting advertisements in posts. These metrics assess detection effectiveness for disclosed and undisclosed ads, including accuracy, precision, recall, and F1 score. We provide a detailed description of each metric in section~\ref{sec:results}, framing them within the context of the experiments they inform and support.

\paragraph{Instagram Dataset}
\label{sec:datasets}

We rely on real Instagram data to evaluate the synthetic captions and provide examples in the prompts. For that purpose, we use the dataset curated by~\citet{bertaglia2023closing}, including 200k posts by micro and mega influencers based in the US, published from 2011 to 2022. The dataset includes influencers on different content segments, including fashion, beauty, lifestyle, food, and sports.
We use two separate datasets for sponsored content detection: one dedicated to disclosed ads and the other for undisclosed ads. We employ the dataset introduced by~\citet{kimDiscoveringUndisclosedPaid2021b} for disclosed ads. This dataset includes 1.6 million posts mentioning brands published from 2013 to 2019 by 38k influencers across various industries. We sample a subset of 2k posts balanced by class and use it as a test set: 1k posts with sponsored content and the remaining 1k with non-sponsored. For undisclosed ads, we use the annotated subset from the dataset by~\citet{bertaglia2023closing}. This subset consists of 804 posts with undisclosed ads, labelled by annotators with expertise in influencer marketing research.
While Instagram's terms of service prevent direct data sharing, we share the shortcodes of all example posts and the specific samples used for evaluation. Researchers can then retrieve the data using these shortcodes.

\section{Empirical Observations}
\label{sec:results}

This section presents a comprehensive analysis of synthetic captions, focusing on their intrinsic characteristics, content, and how they compare against real captions. We analyse textual elements, investigate their semantic content through embeddings, and examine network features. Additionally, we evaluate the performance of these synthetic captions in detecting sponsored content and explore their potential to augment real data.

\paragraph{Experimental Setup}
We generated all synthetic posts using the \textit{gpt-3.5-turbo} model through OpenAI's API~\cite{openai-api}. For each prompt strategy described in section~\ref{sec:methodology}, we produced 1,000 captions, equally divided between sponsored and non-sponsored posts. In each API request, we targeted the generation of 15 captions. However, the output varied, often yielding fewer than 15 captions. To ensure output consistency and format, we used the API's function calling feature~\cite{openai-func} to format outputs into JSON. This approach helped us filter out and discard captions that were not correctly formatted in JSON, as these were likely to contain irrelevant content or not follow the instructions. We also tracked a \textit{success rate} for each prompt strategy, reflecting the proportion of captions correctly parsed without issues.

Due to the synthetic posts' quantity (500 sponsored and 500 non-sponsored for each prompt setting), comparing them directly to the entire real dataset was impractical. This limitation was particularly evident in features that scale with post volume, like the number of hashtags. Consequently, we opted for bootstrapped sampling to create a comparable real dataset sample. We repeatedly sampled 1,000 real posts a hundred times, replacing each sample to minimise selection bias. We then averaged the metrics across these samples to produce a distribution for each metric. By calculating a 95\% confidence interval from these distributions, we confirmed the stability and representativeness of our averaged metrics. All metrics consistently fell within this confidence interval. We used a fixed sample across all experiments for qualitative analyses where bootstrapped sampling was unsuitable.

\paragraph*{Selecting the model temperature}
Our first experiment aims to identify the optimal temperature setting for the model. We used the base prompt without including examples to isolate the temperature's impact. We evaluated the model at temperature settings of 0, 0.25, 0.5, 0.7, and 1. We noted that settings above 1 led to undesirable outcomes, such as hallucinations and repetitive outputs, including irrelevant code snippets. We first assessed the model's performance based on its success rate, defined as the proportion of responses correctly formatted as JSON files without errors. A low success rate implied improperly formatted outputs and invalid content, making the prompt inefficient for scalable use. Additionally, we examined the uniqueness of captions and their overlap with hashtags from real data.

Table~\ref{tab:temperatures} reveals a clear trend: the uniqueness of captions increased with temperature. A setting of 0 produced only 42.3\% unique captions, whereas a temperature of 1 achieved 100\% uniqueness. Conversely, the success rate diminished with higher temperatures, dropping to 74.8\% at a temperature of 1. We also observed that lower temperatures resulted in less overlap of hashtags with the real dataset. However, there were no significant differences in other metrics. Crucially, we found that only temperatures of 0.7 and 1 successfully generated captions with user tags overlapping real data, indicating their ability to produce existing usernames. Given the better success rate at 0.7 compared to 1, we selected a temperature setting of 0.7 for all subsequent experiments.

\begin{table*}
\centering
\caption{Impact of gpt-3.5-turbo's temperature parameter on synthetic dataset characteristics.}
\label{tab:temperatures}
\begin{tabular}{lrrrrr}
\toprule
\bfseries Temperature & \bfseries 0 & \bfseries 0.25 & \bfseries 0.5 & \bfseries 0.7 & \bfseries 1 \\
\midrule
\bfseries Unique Captions (\%) & 42.32 & 92.48 & 97.51 & 99.41 & 100.00 \\
\bfseries Success Rate (\%) & 100.00 & 95.24 & 87.74 & 83.81 & 74.83 \\
\bfseries Hashtag Overlap (\%) & 0.46 & 0.98 & 0.98 & 0.95 & 0.81 \\
\bfseries User Tag Overlap (\%) & 0.00 & 0.00 & 0.00 & 0.07 & 0.07 \\
\bottomrule
\end{tabular}
\end{table*}

\subsection{Characterising Synthetic Captions}
Having established the ideal temperature setting based on unique captions and user tag overlap, we focus on a more in-depth understanding of the generated synthetic data. The various categories of metrics provide an overview of different aspects of the captions' quality. In this subsection, we will analyse the characteristics of the captions generated by the different prompt strategies and compare them to real data. 

\paragraph*{Caption Composition Metrics}
This set of metrics represents the structure and content diversity in the captions. We can understand the nuanced patterns and deviations in synthetic captions by examining average caption lengths, standard deviation, and vocabulary range -- including features inherent to Instagram content, such as hashtags, username mentions, and emojis. Table~\ref{tab:composition_metrics} presents the statistics of caption length and number of hashtags, user tags, and emojis for the captions from each experiment setting.
The results show that real captions have a distinctively longer length, with nearly 43 words on average -- significantly higher than all synthetic captions. The difference in standard deviation is also noteworthy. The length of real captions varies greatly, with a standard deviation of 52.12, while synthetic captions have a more uniform size. This variation is due to the outlier captions with a very high number of tokens. The longest real caption has 483 tokens, while the longest synthetic one out of all settings (with the Imitation strategy) has only 270. 

\begin{table*}
\centering
\caption{Impact of different prompt-engineering strategies on the characteristics of the resulting synthetic dataset; values within brackets represent the number of unique entities.}
\label{tab:composition_metrics}
\begin{tabular}{lcccc}
\toprule
 & \bfseries Caption Length & \bfseries Hashtags & \bfseries User Tags & \bfseries Emojis \\
\midrule
\bfseries Base Prompt & 21.92$\pm$10.92 & 0.96$\pm$0.78 (484) & 0.33$\pm$0.48 (293) & 2.04$\pm$1.34 (257) \\
\bfseries Fixed Examples & 33.49$\pm$24.52 & 1.08$\pm$0.82 (591) & 0.53$\pm$1.14 (206) & 2.33$\pm$2.15 (232) \\
\bfseries Random Examples & 28.38$\pm$16.42 & 1.05$\pm$0.83 (685) & 0.56$\pm$0.73 (502) & 2.12$\pm$1.36 (325) \\
\bfseries Imitation & 35.65$\pm$15.65 & 1.17$\pm$0.90 (899) & 0.70$\pm$0.77 (595) & 2.15$\pm$1.44 (355) \\
\bfseries Real & 42.86$\pm$52.12 & 1.97$\pm$3.20 (1348) & 1.38$\pm$1.87 (996) & 1.88$\pm$2.72 (448) \\
\bottomrule
\end{tabular}
\end{table*}

The Imitation strategy achieves the most similar results to the real data, showing it is more capable of mimicking the style of Instagram captions than the other strategies. Hashtags and user tags, which often encapsulate the essence of Instagram captions, are more abundant in real data. On average, real captions have close to three hashtags and two user tags each. Comparatively, the Imitation strategy again leads among the generation strategies. Still, their frequency on synthetic captions is much lower. Emojis are slightly more prevalent in synthetic captions. Emojis are single tokens and are easier to generate than tags, which have a deeper meaning and require semantic understanding.
The number of unique tags and emojis shows that synthetic captions have considerably less variety than real ones. These results underscore that while the generated synthetic data capture characteristics of real captions, there is a significant gap in content depth and variety.

\paragraph*{Content-specific Metrics}
While the caption composition metrics offer an overview of the distribution of textual elements, they do not describe the actual content within those captions. This subsection further analyses the captions to investigate what they express and how they compare with real ones. We first analyse the overlap between tags and n-grams (ranging from 1 to 3) in synthetic and real captions. For tags, we calculate the proportion of hashtags and user tags from the synthetic data also present in real captions. For n-grams, we use Jaccard similarity for a more detailed comparison. This metric calculates how many n-grams from our synthetic captions are also present in the real ones relative to all the unique n-grams in both sets combined. This method gives a proportion that reflects the similarities and differences between the two datasets, offering a more balanced view than just looking at direct overlaps. Table~\ref{tab:content_metrics} presents the results.

\begin{table*}
\centering
\caption{Content overlap between real data and synthetic datasets generated with different prompting strategies.}
\label{tab:content_metrics}
\begin{tabular}{lrrrr}
\toprule
 & \bfseries Base Prompt & \bfseries Fixed Examples & \bfseries Random Examples & \bfseries Imitation \\
\midrule
\bfseries Hashtag Overlap (\%) & 0.95 & 1.16 & 2.84 & 3.42 \\
\bfseries User Tag Overlap (\%) & 0.07 & 1.12 & 4.52 & 4.05 \\
\bfseries 1-gram Sim. (\%) & 12.72 & 15.16 & 19.75 & 21.24 \\
\bfseries 2-gram Sim. (\%) & 4.70 & 6.01 & 7.36 & 8.36 \\
\bfseries 3-gram Sim. (\%) & 0.99 & 1.71 & 2.23 & 2.48 \\
\bottomrule
\end{tabular}
\end{table*}

The results indicate distinct trends across different synthetic caption strategies. The Imitation method consistently emerges as the closest to real captions, yet its overlaps remain low. Hashtag Overlap peaks at 3.42\%, and User Tag Overlap at 4.05\%  for this strategy, highlighting that the synthetic captions tend to contain tags that do not occur in real data. The 1-gram similarity indicates that about a fifth of individual words from Imitation captions align with real ones. However, when we focus on 2-grams and 3-grams, this percentage drops sharply to 8.36\% and 2.48\%, respectively. This result underscores a challenge: synthetic methods can mimic basic style and content, but capturing nuanced, multi-word expressions and context-rich tags prevalent in real captions remains challenging. To investigate the differences in content more deeply, we analyse both datasets' most common tags and n-grams.

We began by extracting the top one hundred entities from each dataset. Comparing the generated captions to the real ones, we found that their overlaps are generally low. The Random Examples method led in hashtag overlap at just 9\%. Meanwhile, the Fixed Examples reached an 8\% overlap for user tags. When examining n-grams, the Imitation method stood out. It achieved a 54\% overlap for 1-grams, suggesting that over half of the common words in synthetic captions mirrored those in real ones. Yet, overlap percentages drop sharply for multi-word patterns of 2-grams and 3-grams, peaking only at 12\% and 7\%, respectively. This trend reiterates our prior observations: although synthetic captions can capture individual elements of Instagram text, their frequency and combination patterns differ from real data.

We then identified the most frequent elements in real captions missing from the synthetic data and vice versa. We observe that synthetic captions often contain tags that could pass as Instagram vocabulary but do not truly reflect the language used on the platform. Tags such as \#throwback, \#skincareobsessed, \#haircareproducts, and \#coffeelover seem appropriate yet are rarely used in real captions. Moreover, we noted numerous generic brand names in synthetic captions when examining user tags. Tags like @foodieheaven, @proteinshakebrand, and @activelifestyleapparel sound plausible and could even be genuine brands. However, real brand names and specific hashtags used by influencers to uniquely tag their content were frequently missing from the synthetic data. Conversely, while this discrepancy highlights a gap in the fidelity of our synthetic data to real Instagram content, it also points to a positive aspect regarding privacy and data leakage. The absence of actual brand names from the synthetic dataset suggests a reduced risk of inadvertently disclosing real-world data, indicating an inherent advantage in privacy preservation.

The 3-gram analysis further illustrated the characteristics of the synthetic dataset. We identified unusual phrases such as \enquote{love steve jobs} and references to potentially harmful ads such as \enquote{nordic spirit nicotine}. Furthermore, synthetic captions tend to use stereotypical language associated with influencers, with 3-grams such as \enquote{new skincare routine} and \enquote{sunday brunch goals}. On the other hand, the synthetic dataset failed to capture frequently used phrases such as \enquote{like post follow}, \enquote{check link bio}, and \enquote{giveaway affiliated sponsored} -- which are generally prevalent on Instagram and commonly found across influencer profiles.

Continuing our investigation, we analysed the use of emojis in the captions. While synthetic posts have a higher average number of emojis, they lack diversity. The range of unique emojis in the synthetic data varied from 257 in Base Prompt to a peak of 355 in Imitation, while the real data has 448. Upon analysing the emojis missing in synthetic data, we identified a problematic trend: a pronounced lack of variation in skin tones. Many emojis allow selecting from five distinct skin tones (light, medium light, medium, medium dark, and dark) other than the default yellow one. We calculated the prevalence of each tone across all datasets. Table~\ref{tab:skin_tone} presents the results.

\begin{table*}
\centering
\caption{Distribution of emoji skin tones in each dataset.}
\label{tab:skin_tone}
\begin{tabular}{lrrrrr}
\toprule
 & \bfseries Light & \bfseries Medium-Light & \bfseries Medium & \bfseries Medium-Dark & \bfseries Dark \\
\midrule
\bfseries Base Prompt & 36.00 & 26.00 & 0.00 & 0.00 & 0.00 \\
\bfseries Fixed Examples & 27.00 & 17.00 & 50.00 & 0.00 & 0.00 \\
\bfseries Random Examples & 62.00 & 35.00 & 8.00 & 1.00 & 1.00 \\
\bfseries Imitation & 52.00 & 36.00 & 2.00 & 3.00 & 3.00 \\
\bfseries Real & 235.08 & 155.82 & 26.60 & 22.43 & 25.63 \\
\bottomrule
\end{tabular}
\end{table*}

The results highlight the disparities in the representation of diverse skin tones between synthetic and real captions. Base Prompt lacks medium to dark skin tones, mirroring our previous observations of underrepresented entities in synthetic data. Fixed Examples exhibit an unusual over-representation of the medium category with no medium-dark and dark emojis. Conversely, Real captions are more balanced, with a higher frequency of light and medium-light tones and a consistent distribution across all categories. The Random Examples and Imitation methods are closer to the real representation, although with evident discrepancies. Notably, Imitation is the only synthetic strategy with a distribution across all five tones, even though the numbers for medium-dark and dark are considerably lower. It is worth highlighting the significant frequency of light and medium-light tones in the real dataset, underlining their widespread usage. However, the fact that synthetic captions fail to capture the diversity of emoji skin tones -- a fundamental aspect of identity representation on social platforms -- signals an area where the data generation requires more refinement to mimic genuine human expression accurately. These results highlight the broader challenge that even when synthetic strategies approximate some aspects of real content, they fail to capture nuanced details crucial to representation and inclusivity.

In conclusion, while synthetic data, particularly from the Imitation method, can mimic textual elements of real captions, they still fail to capture the complexity and diversity expressed by Instagram captions. This observation, combined with our previous findings, underscores the critical importance of nuanced content understanding for any synthetic model aimed at generating realistic and culturally sensitive content. It also emphasises the potential pitfalls and biases that might inadvertently be introduced, particularly when the data do not thoroughly represent the intricacies of human language and social platform usage. 

\subsection{Embedding Similarity}
Embedding similarity metrics are essential in assessing the semantic similarity between synthetic and real captions. By representing captions in a high-dimensional space, we can compare their contextual and semantic characteristics. The cosine similarity between embeddings quantifies how close the synthetic captions resemble real ones regarding content and meaning. A higher cosine similarity indicates more significant similarity, indicating that the generated content aligns well with real Instagram captions. Nonetheless, it is essential to emphasise that high similarity does not inherently equate to fidelity or realism in the captions but merely suggests aligned content themes. We also calculate the top-100 recall: For each real caption, we retrieve the 100 most similar captions in the embedding space and then measure what fraction of these are synthetic. A high recall value suggests that many of our synthetic captions frequently appear among the top matches of real ones, underscoring the capability of our generation's methods to emulate real Instagram content. Table~\ref{tab:embedding_similarity} presents the results.

\begin{table}
\centering
\caption{Embedding similarity between real data and synthetic datasets generated with different prompting strategies.}
\label{tab:embedding_similarity}
\begin{tabular}{lrr}
\toprule
 & \bfseries Similarity & \bfseries Top-100 Recall \\
\midrule
\bfseries Base Prompt & 0.83$\pm$0.02 & 0.11 \\
\bfseries Fixed Examples & 0.84$\pm$0.05 & 0.12 \\
\bfseries Random Examples & 0.83$\pm$0.03 & 0.10 \\
\bfseries Imitation & 0.83$\pm$0.03 & 0.10 \\
\bottomrule
\end{tabular}
\end{table}

The cosine similarity values across different prompts are consistently around 0.83. This consistency suggests that the generated captions can mimic the semantic content of real Instagram captions regardless of the prompting strategy. The minor variations in similarity scores among the prompts are not significant. Considering the setup of comparing 1k synthetic to 1k real captions, a random selection of the top 100 most similar captions would result in a 50\% recall. However, our results show a top-100 recall significantly below the random baseline, indicating that synthetic captions seldom rank among the closest matches to the real ones. While semantically similar, synthetic captions lack the nuanced patterns or specific vocabulary often found in authentic Instagram content, as we highlighted previously. These observations underscore the importance of adopting a comprehensive set of metrics for analysis, moving beyond relying on simple cosine similarity.

\subsection{Network-based Metrics}
Instagram posts often refer to other users or posts via hashtags, thus creating a network structure. \citet{kimDiscoveringUndisclosedPaid2021b} find that this is a relevant feature of social media data in the context of ad detection. Thus, when creating synthetic data, it is important that its network features mirror those of real data. We investigate three types of networks: 

\begin{itemize}
 \item Hashtag co-occurrence networks, in which nodes represent hashtags and an edge connects two hashtags if they occur in the same post.
 \item User tags co-occurrence networks, in which nodes represent tagged users and an edge connects two nodes if they occur in the same post. 
 \item Hashtags and user tags co-occurrence networks. These are bipartite networks where the two sets of nodes represent hashtags and user tags. Edges connect hashtags and user tags if they appear in the same post.
\end{itemize}

Table~\ref{tab:network_metrics} presents metrics derived from the networks extracted from the four synthetic datasets compared to the corresponding networks from real data.
\begin{table*}
\centering
\caption{Network metrics for the three classes of networks analysed: hashtag co-occurrence (HT), usertag co-occurrence (UT), and hashtag-user bipartite network (HU). Columns correspond to the four synthetic datasets we analyse and the real dataset. The values reported are averaged over 100 network instances. }
\label{tab:network_metrics}
\begin{tabular}{lrrrrr}
\toprule
 & \bfseries Base Prompt & \bfseries Fixed Examples & \bfseries Random Examples & \bfseries Imitation & \bfseries Real \\
\midrule
\bfseries HU \# Nodes & 484.00 & 591.00 & 685.00 & 899.00 & 1350.22 \\
\bfseries HU \# Edges & 237.00 & 308.00 & 356.00 & 505.00 & 5540.52 \\
\bfseries HU Avg. Clustering Coeff. & 0.02 & 0.05 & 0.08 & 0.12 & 0.74 \\
\bfseries HU Avg. Degree & 0.98 & 1.04 & 1.04 & 1.12 & 8.18 \\
\bfseries HU Assortativity & -0.04 & -0.03 & -0.05 & -0.07 & -0.08 \\
\midrule
\bfseries UT \# Nodes & 293.00 & 206.00 & 502.00 & 598.00 & 996.62 \\
\bfseries UT \# Edges & 6.00 & 56.00 & 139.00 & 167.00 & 1682.58 \\
\bfseries UT Avg. Clustering Coeff. & 0.01 & 0.07 & 0.11 & 0.12 & 0.43 \\
\bfseries UT Avg. Degree & 0.04 & 0.54 & 0.55 & 0.56 & 3.35 \\
\bfseries UT Assortativity & 1.00 & 0.89 & 0.70 & 0.99 & 0.40 \\
\midrule
\bfseries HU \# Nodes & 738.00 & 756.00 & 1144.00 & 1395.00 & 2241.48 \\
\bfseries HU \# Edges & 358.00 & 339.00 & 755.00 & 865.00 & 3171.58 \\
\bfseries HU Avg. Degree & 0.97 & 0.90 & 1.32 & 1.24 & 2.83 \\
\bfseries HU Assortativity & 0.09 & -0.16 & -0.00 & -0.05 & -0.09 \\
\bottomrule
\end{tabular}
\end{table*}
All synthetic networks have significantly fewer nodes (between 30\% and 66\%) and edges (0.3\% to 10\%) compared to those extracted from real data. Moreover, consistent with the different ratios between the number of nodes and edges in these networks, the synthetic networks are much less connected than their real counterparts, even though they were built from a randomly chosen subset of posts of the same size.

\begin{figure}[!ht]
    \centering
    \begin{subfigure}[b]{0.48\columnwidth}
        \centering
        \includegraphics[width=\linewidth]{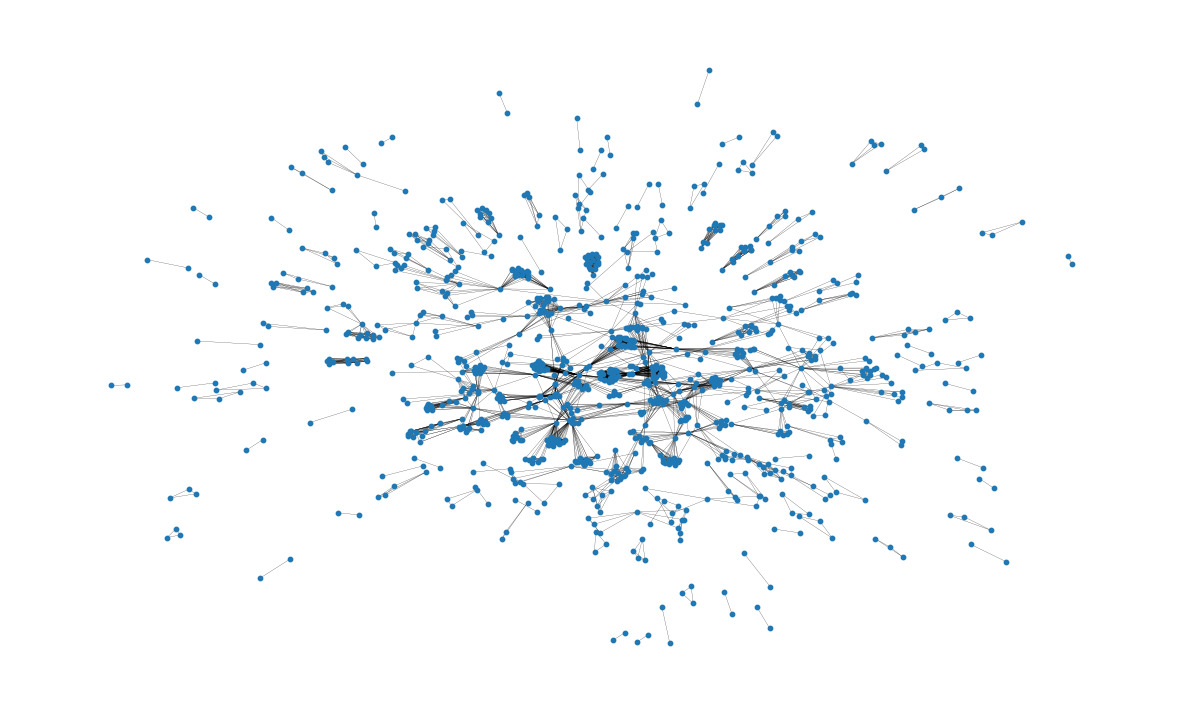}
        \caption{Hashtag co-occurrences in the real data sample.}
        \label{fig:real}
    \end{subfigure}
    \hfill
    \begin{subfigure}[b]{0.48\columnwidth}
        \centering
        \includegraphics[width=\linewidth]{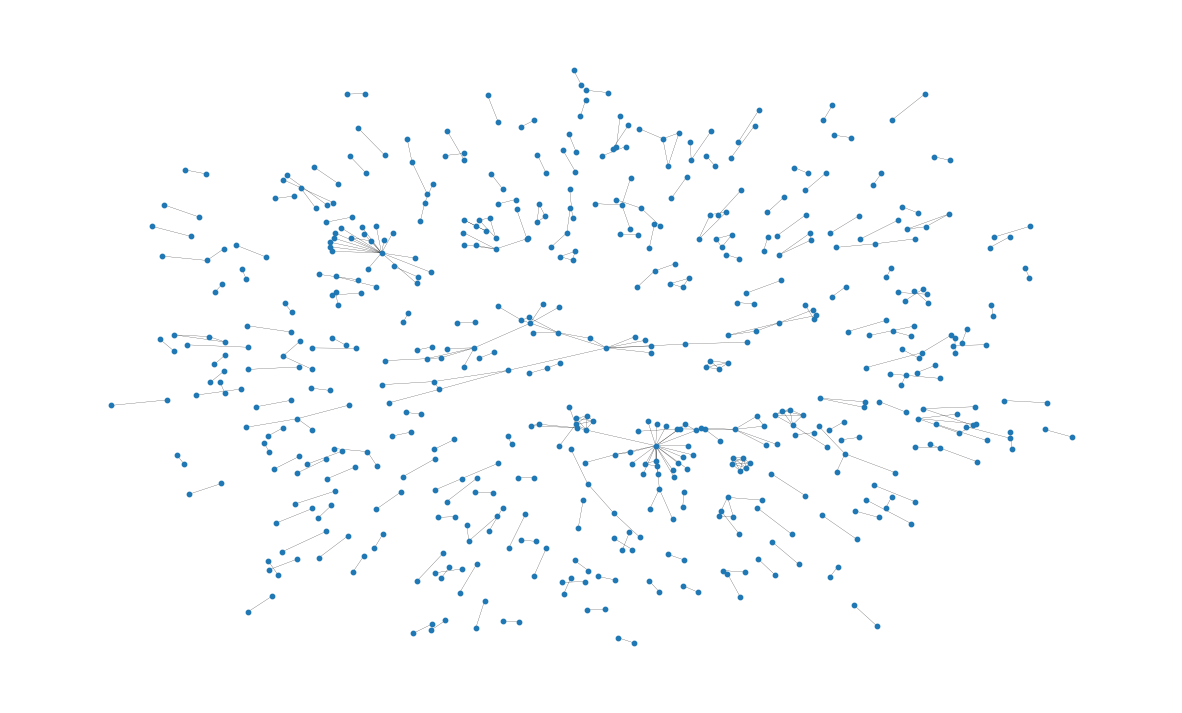}
        \caption{Hashtag co-occurrences in the Imitation synthetic dataset.}
        \label{fig:imitation}
    \end{subfigure}
    \caption{Comparative visualisations of hashtag co-occurrence networks: real vs. synthetic data.}
    \label{fig:combined}
\end{figure}

In addition, the clustering coefficient and degree assortativity of the synthetic networks are significantly lower than those of real ones. This result suggests that synthetic networks are more prone to star-like structures, where few hashtags have very high degrees and hashtags of low degrees are not necessarily connected to each other. It is another way to quantify the effects of chatGPT's preference for specific stereotypical hashtags that are thus overrepresented in the data at the cost of diversity. Figure~\ref{fig:combined} illustrates these features.

Qualitative analysis of the connected components and visualisations of the top 100 most frequent hashtags with labels show that the real data is more likely to contain specific niches that are marked by the use of the same hashtags. Synthetic data exhibits groupings, but they are less clearly delineated in topic, and the topics are less specific. For example, \#FitnessGoals and \#fitnessjourney are unconnected. 
Furthermore, the topics in synthetic data stay quite generic. For example, while real hashtags often mention specific places or events, synthetic hashtags do this rarely. However, the Imitation prompt strategy mitigates this effect to some degree.

\subsection{Downstream Task Performance}
In this study, we evaluate the utility of our synthetic Instagram data in improving the detection of sponsored content, with a specific focus on undisclosed ads. This task is crucial for ensuring transparency, ethical advertising on social media platforms, and protecting consumer interests. We train Logistic Regression models to classify real Instagram data as sponsored or non-sponsored, using the default \textit{scikit-learn} implementation~\cite{scikit-learn}. We chose a simple machine learning approach, employing TF-IDF features from unigrams, and avoided hyperparameter tuning to emphasise the impact of data quality over model complexity. While more complex models like BERT~\cite{devlin2019bert} could potentially offer higher performance, they typically require extensive pretraining on large datasets. In such cases, the influence of the pretraining data may overshadow the impact of the smaller fine-tuning dataset. Moreover, the results from \citet{bertaglia2023closing} showed that a logistic regression model outperformed BERT on undisclosed ad detection. Therefore, our focus remains on assessing the comparative effectiveness of training data rather than pursuing the highest accuracy with advanced model architectures.

We benchmarked our synthetic data's utility by comparing models trained exclusively on real data against those trained on synthetic data. We aimed to assess whether synthetic data, alone or combined with real data, could enhance the model performance in detecting sponsored content. We used two distinct test sets for evaluation. The first set (T1) comprised 1000 disclosed sponsored posts and 1000 non-sponsored posts, sourced from~\citet{kimDiscoveringUndisclosedPaid2021b}. The second set (T2) is the manually annotated subset of the dataset by~\citet{bertaglia2023closing} and includes exclusively undisclosed sponsored posts. We report precision, recall, and F1 scores for T1 and accuracy for T2, as it contains only a single class (undisclosed ads). Table~\ref{tab:downstream_task} presents the performance of models trained on each dataset.

\begin{table}
\centering
\caption{Performance of the logistic regression model trained on different datasets, four synthetically generated and one real. The task is to classify Instagram captions as sponsored or non-sponsored. We used two testing sets: T1 with an equal number of sponsored/non-sponsored posts (from~\citet{kimDiscoveringUndisclosedPaid2021b}) and T2 containing only undisclosed sponsored posts (from~\citet{bertaglia2023closing}). Precision (P), recall (R) and F1 score are reported on T1, and accuracy (Acc) on T2.}
\label{tab:downstream_task}
\begin{tabular}{lcccc}
\hline
                          & \multicolumn{3}{c}{\bfseries T1}                                        &  \bfseries T2             \\ \hline
\bfseries Training Data   & \bfseries P & \bfseries R & \multicolumn{1}{c|}{\bfseries F1} & \bfseries Acc. \\ \hline
\bfseries Base Prompt     & 0.53        & 0.79        & \multicolumn{1}{c|}{0.64}         & 0.74           \\
\bfseries Fixed Examples  & 0.65        & 0.68        & \multicolumn{1}{c|}{0.66}         & 0.40           \\
\bfseries Random Examples & 0.63        & 0.82        & \multicolumn{1}{c|}{0.71}         & 0.51           \\
\bfseries Imitation       & 0.72        & 0.69        & \multicolumn{1}{c|}{0.71}         & 0.23           \\
\bfseries Real            & 0.66        & 0.88        & \multicolumn{1}{c|}{0.76}         & 0.49           \\ \hline
\end{tabular}
\end{table}

Real data achieves the highest recall and F1 for disclosed ads; both Random examples and Imitation datasets perform comparably, while Base Prompt and Fixed Examples are significantly worse. Base Prompt dataset achieves a high recall, showing that the model trained on that dataset correctly detects most sponsored posts. However, its precision is very low (0.53), suggesting it biases the model toward overclassifying captions as ads. While this might be advantageous in ensuring minimal missed sponsored content, it simultaneously risks increasing the false positive rate, considering regular posts as ads. This trade-off is also noticeable with the Random Examples dataset and, to a lesser extent, with the real data. Contrarily, the Imitation dataset obtains a balance between the two metrics.

However, the model performance for undisclosed ads is significantly different. The generally low metrics (0.49 accuracy with real data and 0.23 with Imitation) highlight the challenges of detecting undisclosed advertisements. Yet, the Base Prompt surprisingly outperforms all settings, with an accuracy of 0.74, while being the least similar to actual Instagram captions. This result indicates that \enquote{synthetic} or unrealistic characteristics inherent to the Base Prompt's captions could be useful in detecting undisclosed ads. This observation suggests a potential conflict in the metrics used to measure the fidelity of synthetic data:  Imitation, despite leading in textual and content similarity to real posts, performs poorly in the downstream task. In contrast, the Base Prompt, diverging in textual fidelity, displays a significantly superior performance. 

To further investigate how dataset characteristics influence model performance, we analysed the vocabulary overlap between the training datasets and sponsored captions from the test sets. Specifically, we calculated the percentage of common tokens between them. This measure shows how each dataset incorporates the lexicon indicative of sponsored content. By analysing these overlaps, we aim to identify lexical biases that might influence the model's performance in detecting sponsored content. Table~\ref{tab:ad_detection_overlap} presents the overlap percentage across all datasets.

\begin{table}
\centering
\caption{Overlap percentage between unigrams from each dataset and the test sets. \emph{Disclosed} refers to only the sponsored posts from T1. \emph{Undisclosed} refers to all posts in T2.}
\label{tab:ad_detection_overlap}
\begin{tabular}{lrr}
\toprule
 \bfseries Dataset & \bfseries Disclosed & \bfseries Undisclosed \\
\midrule
\bfseries Base Prompt & 56.67 & 48.46 \\
\bfseries Fixed Examples & 55.02 & 48.07 \\
\bfseries Random Examples & 52.77 & 45.98 \\
\bfseries Imitation & 50.25 & 44.61 \\
\bfseries Real & 44.24 & 40.22 \\
\bottomrule
\end{tabular}
\end{table}

The real data has the lowest overlap with both disclosed and undisclosed test sets, suggesting it might be the least representative of the vocabulary found in the test captions. On the other hand, Base Prompt has the highest overlap, corroborating our earlier observations about the downstream task performance and hinting at an inherent vocabulary bias. Yet, while showing a considerable overlap with the undisclosed ads, the other synthetic datasets do not mirror the high accuracy exhibited by the Base Prompt. The discrepancy in vocabulary diversity could significantly contribute to this gap: Base Prompt's data has the fewest unique tokens. This lack of diversity might simplify the learning process for the model, enabling it to learn associations between frequent tokens common to sponsored content. 

In this scenario, \enquote{realistic} captions could be undesirable. Their rich lexical diversity adds complexity to the task, inflating the overall vocabulary and feature set. While this point might seem disadvantageous on the surface, diverse data potentially results in models with greater generalisation capabilities that can recognise a wider array of features and better capture the characteristics of sponsored content. Base Prompt's success, in contrast, may rely on token coincidence rather than a profound understanding. Therefore, improvements in downstream task metrics do not necessarily imply high-quality synthetic data; a superficial spike in performance might mask underlying dataset shortcomings. Considering a broader set of metrics and holistically evaluating the synthetic captions is essential to assess quality effectively.

Our results so far relied on individual datasets for training. Our next experiment analyses the benefits of integrating synthetic with real data. The synthetic datasets, while limited, contain a more focused vocabulary. In contrast, real data has higher lexical diversity and offers robustness and potential for broader generalisation. By merging these datasets, we hypothesise that the combined training data can encapsulate characteristics from both, thus improving the model's performance in detecting sponsored content. To test this hypothesis, we merged each synthetic dataset with real captions and repeated the downstream task experiments. We combined each synthetic dataset (with 1k captions in total) with 1k real data samples. To control for the influence of the random sample, we repeated this process 100 times, as detailed in Section~\ref{sec:methodology}. To ensure the difference in performance was not only due to the increase in training data, we also trained a model on 2000 real captions, applying the same bootstrapping approach. Table~\ref{tab:ad_detection_aug} presents the performance metrics of models trained on the augmented datasets.

\begin{table}
\centering
\caption{Performance of the logistic regression model trained on combined datasets of 1000 synthetic and 1000 real Instagram captions. We used the same testing sets: T1 with an equal number of sponsored/non-sponsored posts and T2 containing only undisclosed sponsored posts. Precision (P), recall (R) and F1 score are reported on T1, and accuracy (Acc) on T2.}
\label{tab:ad_detection_aug}
\begin{tabular}{lcccc}
\hline
                          & \multicolumn{3}{c}{\bfseries T1}                                        &  \bfseries T2             \\ \hline
\bfseries Training Data   & \bfseries P & \bfseries R & \multicolumn{1}{c|}{\bfseries F1} & \bfseries Acc. \\ \hline
\bfseries Real (1k) + Base (1k) & 0.63 & 0.90 & \multicolumn{1}{c|}{0.74} & 0.57 \\
\bfseries Real (1k) + Fixed Ex. (1k) & 0.68 & 0.89 & \multicolumn{1}{c|}{0.77} & 0.47 \\
\bfseries Real (1k) + Rnd Ex. (1k) & 0.66 & 0.89 & \multicolumn{1}{c|}{0.76} & 0.48 \\
\bfseries Real (1k) + Imitation (1k) & 0.70 & 0.88 & \multicolumn{1}{c|}{0.78} & 0.38 \\
\bfseries Only Real (2k) & 0.69 & 0.89 & \multicolumn{1}{c|}{0.78} & 0.42 \\
\bottomrule
\end{tabular}
\end{table}

The combined datasets consistently improved model performance for disclosed ads across all settings. Each merged dataset surpassed the performance of its single dataset counterpart. Notably, the values for precision and recall are significantly more balanced; the combined datasets noticeably improve the precision scores. While the model trained exclusively on real data still outperformed others, it is important to note that the performance gains are not only due to increased training data volume. Doubling the size of the real dataset did not correspond to major improvements in results. This observation suggests that combining synthetic and real datasets provides a nuanced variety and feature density, providing models with a more comprehensive grasp of the task.

The undisclosed ad accuracy metrics show a more stable variation range than individual datasets. Yet, introducing real captions to the Base Prompt severely decreased its accuracy, dropping from 0.74 to 0.57. This drop can be potentially attributed to a \textit{vocabulary dilution effect}: the specific, and perhaps overfitted, vocabulary of the Base Prompt becomes slightly diluted when combined with the diverse lexicon of the real data. Such dilution can challenge the model, demanding more intricate discernment rather than relying on a token coincidence, leading to the observed drop in accuracy.

Intriguingly, even the accuracy of the model trained only on real data declined. This result emphasises that merely expanding the training data does not guarantee better performance. Instead, the combination of diverse vocabularies and contexts in the augmented datasets may introduce nuances that, while beneficial for detecting disclosed ads, might pose challenges for maintaining consistently high accuracy for undisclosed ads. This disparity underscores an inherent dichotomy between undisclosed and disclosed ads, suggesting that models struggle to generalise across both categories based only on the training data provided.

The results, especially in the context of the Base Prompt, emphasise the importance of finding a balance: while synthetic data can offer specific advantages, over-reliance on them without the richness of real data can expose vulnerabilities in model performance. In conclusion, data augmentation can help improve sponsored content detection, with synthetic datasets complementing and amplifying real data's inherent qualities.

\section{Summary and Discussions}
This paper investigates the feasibility of creating synthetic Instagram captions using LLMs to augment or replace real datasets for scientific purposes. We propose a set of metrics and an evaluation methodology to assess the quality of synthetic social media data. Our experiments use gpt-3.5-turbo, but our methodology can be extended to any LLM. Our investigation shows that, at the individual post level, chatGPT outputs authentic Instagram posts, generating brand names and hashtags with convincing names that are semantically connected to the post itself. It also includes emojis that are appropriately placed and semantically coherent. However, the generated captions use stereotypical language associated with influencers, lacking the depth of real content. Additionally, we highlight that relying on simple metrics such as cosine similarity can be misleading: despite high similarity values, the closest matches to real captions are still other real Instagram posts. Importantly, while the synthetic captions can mimic influencer-style language, they frequently overlook crucial and commonly used phrases typical of authentic Instagram interactions, such as \enquote{like post follow}, \enquote{check link bio}, and \enquote{giveaway affiliated sponsored}. Including these phrases through post-processing methods could significantly enhance the fidelity of the synthetic data.

We experimented with four prompting strategies. The Imitation prompt strategy generates captions with better \emph{diversity} in length, unique hashtags and user tags, and emojis, thereby indicating that this strategy generates more representative captions. Furthermore, we observe that Imitation prompting results in captions with better \emph{distribution}, the highest overlap with the real dataset regarding hashtags, user tags, unigrams, bigrams, and trigrams. Yet, we identified a bias in emoji skin tones across all strategies, with only the imitation strategy offering a sparse representation of darker tones. From a network perspective, hashtag and user tag relationships in synthetic captions lacked the depth and connectivity seen in real captions. This result suggests real users adapt and evolve their tagging practices based on peer interactions, an element absent in independent, batch-generated, synthetic captions.

Our valuation of machine learning models trained on chatGPT-generated captions revealed mixed results. While the model effectively identified disclosed advertisements in Instagram captions, its performance on undisclosed ads was notably worse. Still, synthetic data led to a higher accuracy than models trained only on real data. We identify three central challenges in chatGPT-generated captions: diversity, distribution, and connectivity. Diverse and representative outputs are crucial for reliable research. Experimenting with a broader range of prompts could lead to more realistic results; task-specific prompts, especially ones instructing chatGPT to include relevant hashtags, could help improve diversity. Moreover, improving the interconnectedness of hashtags and user tags using post-processing methods, such as training models to predict hashtags and network connections, is essential.

A key takeaway from our experiments is that the goals of creating synthetic datasets mirroring real ones (\textit{fidelity)} and improving performance on downstream tasks (\textit{utility}) are conflicting. For instance, our attempt to identify undisclosed sponsored posts was most successful when training on synthetic data that diverged significantly from actual Instagram captions. These results suggest that such synthetic data might capture unique traits that machine learning models can exploit. Therefore, selecting an appropriate \enquote{utility measure} is vital when designing synthetic datasets. Valuing only similarity against real data can be limiting, especially when the real dataset lacks complex variations or is related to low-frequency or hard-to-capture phenomena -- such as undisclosed advertising or online toxicity. Moreover, our findings emphasise that while prompt design is essential, it is not enough for generating high-quality synthetic datasets. Post-processing methods to enhance the data's diversity, distribution, and connectivity are a promising step toward better synthetic data.

While we used a specific LLM in this work, our contributions are not tied to its current capabilities. Specifically, we proposed a set of metrics that are relevant for measuring the quality and utility of synthetic social media datasets that can be applied independently of the LLM model used and partially independent of the target downstream task. As such, in a fast-evolving field, our measurements can show which LLMs perform better and in what metrics for generating synthetic datasets for research in computational social science. 

\paragraph{Limitations and Future Work}
This work can be extended in various directions, including more experimentation with prompt engineering, testing other LLMs, developing post-processing techniques, and exploring a more comprehensive range of downstream tasks of varying complexity. We believe that the creation of useful synthetic data is a promising direction for social media research and could offset the limitations brought by the closing of social media platform APIs to researchers, scarcity of data for rarely detected instances (such as coordination operations, disinformation campaigns) and the overall cost of data collection. 

\paragraph*{Broader Perspective and Ethics}
\citet{bisbee2023artificially} show that closed-source LLMs exhibited systematic biases when used to generate synthetic survey responses. Analogously, our results show that captions generated by chatGPT tend to be stereotypical and misrepresent real data. Therefore, it is essential to work on mitigating these issues to use LLM-generated data for social media research reliably. Our work addresses these limitations critically, and we encourage researchers to investigate post-processing techniques to improve diversity in synthetic data. 

We show that LLMs can help improve sponsored content detection. However, they can also be used to produce harmful content. In our research context, LLMs could generate artificial persuasive sponsored content impersonating influencers on platforms like Instagram and be used to promote, for instance, harmful and illicit products. Nevertheless, disseminating such content through official, verified influencer accounts would be challenging, which minimises widespread adverse impacts. 

We do not perform user profiling or violate users' privacy data. Our focus remains on public figures and content shared openly on Instagram.

\bibliography{main}

\subsection{Paper Checklist to be included in your paper}

\begin{enumerate}

\item For most authors...
\begin{enumerate}
    \item  Would answering this research question advance science without violating social contracts, such as violating privacy norms, perpetuating unfair profiling, exacerbating the socio-economic divide, or implying disrespect to societies or cultures?
    \answerYes{Yes, see sections Broader Perspective and Ethics and Limitations and Future Work}
  \item Do your main claims in the abstract and introduction accurately reflect the paper's contributions and scope?
    \answerYes{Yes}
   \item Do you clarify how the proposed methodological approach is appropriate for the claims made? 
    \answerYes{Yes, see sections Methodology and Experimental Setup}
   \item Do you clarify what are possible artifacts in the data used, given population-specific distributions?
    \answerYes{Yes, we discuss this throughout the paper, especially in section Content Analysis}
  \item Did you describe the limitations of your work?
    \answerYes{Yes, see section Limitations and Future Work}
  \item Did you discuss any potential negative societal impacts of your work?
    \answerYes{Yes, see sections Broader Perspective and Ethics and Limitations and Future Work}
      \item Did you discuss any potential misuse of your work?
    \answerYes{Yes, see sections Broader Perspective and Ethics and Limitations and Future Work}
    \item Did you describe steps taken to prevent or mitigate potential negative outcomes of the research, such as data and model documentation, data anonymization, responsible release, access control, and the reproducibility of findings?
    \answerYes{Yes, see sections Broader Perspective and Ethics and Limitations and Future Work}
  \item Have you read the ethics review guidelines and ensured that your paper conforms to them?
    \answerYes{Yes}
\end{enumerate}

\item Additionally, if your study involves hypotheses testing...
\begin{enumerate}
  \item Did you clearly state the assumptions underlying all theoretical results?
    \answerNA{N/A}
  \item Have you provided justifications for all theoretical results?
    \answerNA{N/A}
  \item Did you discuss competing hypotheses or theories that might challenge or complement your theoretical results?
    \answerNA{N/A}
  \item Have you considered alternative mechanisms or explanations that might account for the same outcomes observed in your study?
    \answerNA{N/A}
  \item Did you address potential biases or limitations in your theoretical framework?
    \answerNA{N/A}
  \item Have you related your theoretical results to the existing literature in social science?
    \answerNA{N/A}
  \item Did you discuss the implications of your theoretical results for policy, practice, or further research in the social science domain?
    \answerNA{N/A}
\end{enumerate}

\item Additionally, if you are including theoretical proofs...
\begin{enumerate}
  \item Did you state the full set of assumptions of all theoretical results?
    \answerNA{N/A}
	\item Did you include complete proofs of all theoretical results?
    \answerNA{N/A}
\end{enumerate}

\item Additionally, if you ran machine learning experiments...
\begin{enumerate}
  \item Did you include the code, data, and instructions needed to reproduce the main experimental results (either in the supplemental material or as a URL)?
    \answerYes{Yes, the code and instructions for reproducing the experiments are available on GitHub~\cite{github-repo}. We cannot share Instagram data directly, but we share list of ids of all posts used in the experiments}
  \item Did you specify all the training details (e.g., data splits, hyperparameters, how they were chosen)?
    \answerYes{Yes, see section Downstream Task Performance}
     \item Did you report error bars (e.g., with respect to the random seed after running experiments multiple times)?
    \answerNA{N/A}
	\item Did you include the total amount of compute and the type of resources used (e.g., type of GPUs, internal cluster, or cloud provider)?
    \answerNA{N/A, our machine learning experiments use a simple Logistic Regression model trained on small datasets and do not require intense computing}
     \item Do you justify how the proposed evaluation is sufficient and appropriate to the claims made? 
    \answerYes{Yes, see section Downstream Task Performance}
     \item Do you discuss what is ``the cost`` of misclassification and fault (in)tolerance?
    \answerYes{Yes, see section Downstream Task Performance}
  
\end{enumerate}

\item Additionally, if you are using existing assets (e.g., code, data, models) or curating/releasing new assets, \textbf{without compromising anonymity}...
\begin{enumerate}
  \item If your work uses existing assets, did you cite the creators?
    \answerYes{Yes, see section Instagram Dataset}
  \item Did you mention the license of the assets?
    \answerYes{Yes, we mention the limitations of sharing Instagram data in section Instagram Dataset}
  \item Did you include any new assets in the supplemental material or as a URL?
    \answerNA{N/A}
  \item Did you discuss whether and how consent was obtained from people whose data you're using/curating?
    \answerNA{N/A}
  \item Did you discuss whether the data you are using/curating contains personally identifiable information or offensive content?
    \answerYes{Yes, see section Broader Perspective and Ethics}
\item If you are curating or releasing new datasets, did you discuss how you intend to make your datasets FAIR (see \citet{fair})?
\answerNA{N/A}
\item If you are curating or releasing new datasets, did you create a Datasheet for the Dataset (see \citet{gebru2021datasheets})? 
\answerNA{N/A}
\end{enumerate}

\item Additionally, if you used crowdsourcing or conducted research with human subjects, \textbf{without compromising anonymity}...
\begin{enumerate}
  \item Did you include the full text of instructions given to participants and screenshots?
    \answerNA{N/A}
  \item Did you describe any potential participant risks, with mentions of Institutional Review Board (IRB) approvals?
    \answerNA{N/A}
  \item Did you include the estimated hourly wage paid to participants and the total amount spent on participant compensation?
    \answerNA{N/A}
   \item Did you discuss how data is stored, shared, and deidentified?
   \answerNA{N/A}
\end{enumerate}

\end{enumerate}



\end{document}